\documentclass[12pt]{iopart}
\usepackage{graphicx}
\begin{document}
\title[ Tidal effects on magnetic gyration]
{
Tidal effects on magnetic gyration of a charged particle 
in Fermi coordinates}
\author{Yasufumi Kojima\footnote[1]{Email:
 \mailto{kojima@theo.phys.sci.hiroshima-u.ac.jp} }
and Kentaro Takami\footnote[2]{Email:
 \mailto{takami@theo.phys.sci.hiroshima-u.ac.jp} }
}
\address{Department of Physics,
Hiroshima University,
Higashi-Hiroshima 739-8526, Japan}
\begin{abstract}
  We examine the gyration motion of a charged particle, 
viewed from a reference observer
falling along the $Z$ axis into a Schwarzschild  black hole.
It is assumed that the magnetic field is constant and 
uniform along the $Z$ axis, and that the particle 
has a circular orbit in the $X$-$Y$ plane far from the gravitational source. 
When the particle as well as the reference observer
approaches the black hole,
its orbit is disrupted by the tidal force. 
The final plunging velocity increases in
the non-relativistic case, but decreases
if the initial circular velocity exceeds a critical value 
of $\sim 0.7c$. 
This toy model suggests that disruption of 
a rapidly rotating star due to  
a velocity-dependent tidal force may be quite different from 
that of a non-relativistic star. 
The model also suggested that collapse of the orbit 
after the disruption is slow in general, so that 
the particle subsequently escapes outside the valid Fermi 
coordinates. 
\end{abstract}
   \maketitle
%
\section{ Introduction }

  A star closely orbiting a black hole
may be disrupted by a tidal force,
most likely resulting in highly luminous flares.
Observation of such flares can thus provide important 
clues to the physics of black holes (e.g, \cite{rees88}).
The high luminosity phase is of short duration, 
making detection rare, but recent X-ray 
observations\cite{KHSHSP04,HGK04} suggest a declining phase 
after the tidal disruption in  a non-active galactic nucleus. 
A numerical simulation of tidal disruption
and subsequent dynamical evolution 
has been carried out using smoothed particle  
hydrodynamics \cite{LMZD93}, in which 
the typical velocity of the fluid elements was mildly relativistic,
less than  0.1$c$.

  Recently, Chicone and Mashhoon \cite{CM04} pointed out
a very interesting phenomenon of tidal forces.
When an ultra-relativistic velocity
is considered,
the tidal force exhibits an unexpected property
contrary to non-relativistic dynamics.
Suppose that an object of finite size falls towards
a black hole along the $Z$ axis.
The object would be stretched 
along the $Z$ axis and compressed along the $X$ and $Y$ axes
under the Newtonian tidal force. 
The tidal force would therefore cause anisotropic  
acceleration and deceleration in  the particle dynamics
near a freely-falling observer. 
That is, from the viewpoint of the observer,  
out-going particles on the $Z$ axis would accelerate,
while out-going particles on the $X$ or $Y$ axes 
would decelerate.
This is a non-relativistic expectation. 
Chicone and Mashhoon  however showed that
when the particle velocity 
exceeds a critical value  $  c/\sqrt{2} \sim 0.7 c$,
the particles would decelerate along the $Z$ axis, 
but accelerate along the $X$ or $Y$ axes.
Chicone, Mashhoon and  Punsly \cite{CMP05} also 
discussed the case of a spinning particle 
orbiting a rotating black hole.
In their one-dimensional model,
the relativistic particles instantaneously escape from
a valid region of the local observer frame.

  Here, we present another concrete model of the tidal effects
on relativistic motion.
We consider motion of a charged particle in a weak magnetic field, 
and Schwarzschild metric. 
Fermi normal coordinates are constructed
around the reference observer falling into a black hole.
The magnetic field is assumed to be uniform along the $Z$ axis
in this frame.  
Far from the gravitational source where the tidal force is negligible,
the orbit is circular in the $X$-$Y$ plane.
As the reference frame approaches the black hole along the $Z$ axis,
the tidal force becomes important. But how does it disturb the circular motion?
In particular, how does the relativistic rotating velocity
affect the result?
The spatial orbit of the particle 
is always bound, so that it is possible to pursue 
the long term evolution in the Fermi system.
This toy problem may deepen our understanding of the
tidal force.

  This paper is organized as follows.
In section 2, the equation of motion of a charged particle in 
Fermi coordinates is discussed.
In section 3, the tidal effect in the non-relativistic limit
is discussed.
In section 4, motion with
relativistic velocity is considered, and numerical results are given.
Section 5 is the discussion.

\section{ Basic equations }

  We consider a reference observer whose world line is geodesic. 
Fermi normal coordinates, $(cT,X^k)$ 
in a neighborhood of the observer
can be constructed such that 
the world line of the observer
is given by the observer's proper time $T $ and $X^k = 0$.
The metric accurate to second order in $X^k$ 
takes the form\cite{mtw}:
\begin{equation}
 g_{00}=-1- ~^F R_{0i0j}X^iX^j+ \cdots , 
\end{equation}
\begin{equation}
 g_{0i}=-\frac{2}{3} ~^F R_{0jik}X^jX^k+ \cdots, 
\end{equation}
\begin{equation}
 g_{ij}=\delta_{ij}-\frac{1}{3} ~^F R_{ikjl}X^kX^l+ \cdots ,
\end{equation}
where  $~^F R_{\alpha \beta \gamma \delta }$
are components of the Riemann tensor along the observer.
  We consider the dynamics of 
a particle with mass $m$ and charge $q$,
labeled  
$X^k(T)$ or $ (X(T), Y(T), Z(T) )$ 
in the Fermi system.
The equations of motion for the charged particle subject to 
the Lorentz force are \cite{CM04}
\begin{eqnarray}
&&
\ddot{X^i}+ \Bigl(
  ~^FR_{0i0j}+2 ~^F R_{ikj0} \dot{X^k}
+ 2 ~^F R_{0kj0} \dot{X^i}  \dot{X^k} 
\nonumber
\\
&&
~~~~~~~
+\frac{2}{3}  ~^F R_{ikjl} \dot{X^k}  \dot{X^l} 
+\frac{2}{3} ~^F R_{0kjl} \dot{X^i} \dot{X^k} \dot{X^l}
\Bigr) X^j 
= \frac{1}{\Gamma^2}\left( A^i- A^0 \dot{X^i} \right) ,
\label{mster}
\end{eqnarray}
where the overdots denote differentiation
with respect to time $cT$, and
\begin{equation}
\fl~~
\Gamma ^{-2} = 1- \dot{X_i} \dot{X^i}
+ ~^F R_{0i0j} X^i X^j 
+\frac{4}{3} ~^F R_{0jik} X^j \dot{X^i} X^k 
+\frac{1}{3} ~^F R_{ikjl} \dot{X^i} X^k  \dot{X^j} X^l,
\end{equation}
\begin{equation}
\fl~~
 A^\alpha = \frac{q}{m} F^{\alpha \beta} U_{\beta} .
 \label{maxwell}
\end{equation}
In eq.(\ref{maxwell}), $ F^{\alpha \beta} $
is the Faraday-Maxwell tensor and 
$U^\alpha  = ( \Gamma ,~  \Gamma \dot{X^i} )$ is the four velocity of
the charged particle.
We assume for simplicity that
the electric field is zero and that
the magnetic field is constant along 
the $Z$ direction in the reference observer system.
For motion of a particle 
in a uniform magnetic field $ B $,
the electromagnetic acceleration (\ref{maxwell}) is given by
\begin{equation}
\left(A^0 ,~  A^1,~  A^2,~  A^3 \right)
=\left(0,~  \omega \Gamma  \dot{Y} ,~ 
- \omega \Gamma \dot{X},~ 0  \right),
\end{equation}
where $ \omega =q B/(mc)$ is the cyclotron frequency.

  Using an explicit form of the Riemann tensor in Schwarzschild
spacetime, eq.(\ref{mster}) can be reduced to
\begin{equation}
\fl~~
\ddot{X}+ K X \left(
	   1-2\dot{X}^2+\frac{4}{3}\dot{Y}^2-\frac{2}{3}\dot{Z}^2\right)
-\frac{2}{3} K \dot{X}\left( 5Y\dot{Y}-7Z\dot{Z}\right) 
= + \frac{ \omega}{c \Gamma } \dot{Y} ,
\label{eqx}
\end{equation}
\begin{equation}
\fl~~
\ddot{Y} + K Y \left(
	   1-2\dot{Y}^2+\frac{4}{3}\dot{X}^2-\frac{2}{3}\dot{Z}^2\right)
-\frac{2}{3} K \dot{Y}\left( 5X\dot{X}-7Z\dot{Z}\right) 
 = - \frac{ \omega}{c \Gamma } \dot{X} , 
\label{eqy}
\end{equation}
\begin{equation}
\fl~~
\ddot{Z}- 2K Z \left(
1-2\dot{Z}^2+\frac{1}{3} \dot{X}^2 +\frac{1}{3} \dot{Y}^2 \right)
  -\frac{4}{3} K \dot{Z}\left(X\dot{X}+Y\dot{Y} \right)  = 0 ,
\label{eqz}
\end{equation}
where 
\begin{equation}
K = \frac{GM}{c^2 r^3 },
\label{curv}
\end{equation}
and $r$ is the Schwarzschild radius, 
$G$ the gravitational constant and $M$ the mass.
For a freely falling body
with zero radial velocity at infinity, 
$ r$ can be expressed as a function of proper time $T$
of the reference observer:
$ r = (9GM T^2/2 )^{1/3}$, and therefore
$ K  =  2/(9 c^2 T^2)$.
We consider only the negative region of time $T$,
in which $ T= - \infty $ corresponds to $ r = + \infty $,
and $ T =0 $ corresponds to $ r = 0$.
However, when the spatial range is limited, time is also limited, e.g, 
when the spatial range is limited to $ r \ge r_0$, time is limited to 
$ T \le - \sqrt{  2 r_0 ^3/(9GM) } $. 

  Equation (\ref{eqz}) shows that 
if the particle is initially moving on a $Z=0 $ plane,
(i.e. $ Z=0$ and $\dot{Z} =0$), 
the particle always remains on the plane.
From now on, we ignore motion in the $Z$ direction.

\section{ Non-relativistic limit }

  In this section, we consider the non-relativistic limit of 
eqs.(\ref{eqx}) and (\ref{eqy}), and
assume flat spacetime.
The relevant equations are 
\begin{eqnarray}
&& {\ddot X}  + K X =  +  \frac{\omega}{c}  {\dot Y} , 
\label{cyc.eqn1}
\\
&& {\ddot Y}  + K Y =  - \frac{\omega}{c}  {\dot X} . 
\label{cyc.eqn2}
\end{eqnarray}
The analytic solutions of eqs.(\ref{cyc.eqn1}) and (\ref{cyc.eqn2})
are expressed by Bessel $J_{1/6} $
and Neumann $N_{1/6} $ functions of order $1/6$:
\begin{equation}
 X = 
\sqrt{ 2 \pi \eta} 
\left[  A \cos\left( \eta + \chi-\frac{\pi}{3}  \right)
J_{1/6} (\eta)
       + B \sin\left( \eta + \psi-\frac{\pi}{3}  \right)
N_{1/6} (\eta) 
 \right],
\label{anlx.eqn}
\end{equation}
\begin{equation}
 Y = 
\sqrt{ 2 \pi \eta}
\left[  A \sin\left( \eta + \chi -\frac{\pi}{3}  \right)
J_{1/6} (\eta)
       - B \cos\left( \eta + \psi-\frac{\pi}{3} \right)
N_{1/6} (\eta)
 \right],
\label{anly.eqn}
\end{equation}
where $ \eta = \omega |T|/2$ and  
 $ A, B,  \chi ,  \psi $ are integration constants.
The meaning of these parameters becomes clear  by taking 
the limit as infinity.
For large $|T|$, we have
\begin{eqnarray}
\fl~~
 X & \to & 
  -A \left[ \cos\left( \omega  |T| + \chi +\frac{\pi}{3}\right)
  - \cos \chi  \right]
+ B \left[ \cos\left( \omega  |T| + \psi +\frac{\pi}{3}\right) 
 + \cos \psi  \right],
\\
\fl~~
 Y & \to & 
  -A \left[ \sin\left( \omega  |T|  + \chi +\frac{\pi}{3}\right)  
- \sin\chi  \right]
+ B \left[ \sin\left( \omega  |T|  + \psi +\frac{\pi}{3}\right) 
 + \sin\psi  \right]. 
\end{eqnarray}
The  orbit for large $|T|$ represents a circle with radius
$ ( A^2 + B^2 -2 A B  \cos (\chi- \psi) )^{1/2} $ 
and origin
 $(A \cos \chi +B  \cos  \psi , A \sin \chi +B \sin \psi ). $
Parameterization with  $ A, B,  \chi ,  \psi $ is
awkward for large $|T|$, but useful for small $ |T|$
because two solutions,  referred to as solution A ($ A \ne 0, B=0$) 
and  solution B ($ A =0, B \ne 0$), 
are remarkably  different.
Since $ J_{1/6} (\eta) \sim \eta ^{1/6} $ and
 $ N_{1/6} (\eta) \sim \eta ^{-1/6} $,
for small $ \eta$,  we have 
$ X, Y  \propto A \times ( \eta )^{2/3} $, 
$ \dot{X}, \dot{Y}   \propto A \times ( \eta )^{-1/3} $
in solution A, while 
$ X, Y  \propto B \times ( \eta )^{1/3} $, 
$ \dot{X}, \dot{Y}   \propto B \times ( \eta )^{-2/3} $
in solution B.
Both solutions imply that the orbits converge at 
$ X=Y =0$, with the velocity diverging as $|T|$ tends to zero.
The velocity in solution B
increases rapidly compared with that in solution A.
General solutions are expressed by linear combination of
the two solutions.  

\begin{figure}[ht]
\begin{center}
\includegraphics[scale=0.40]{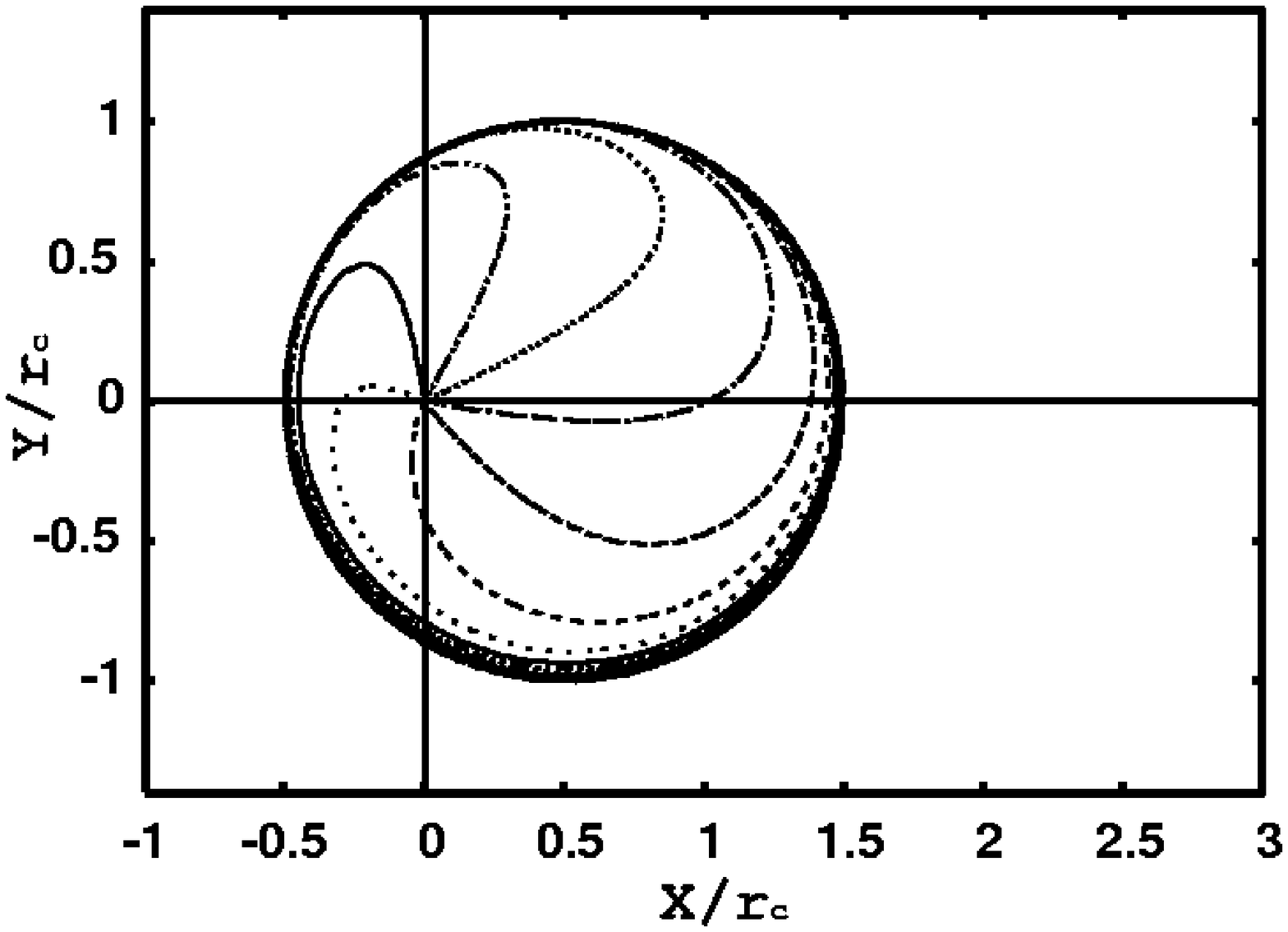}
\includegraphics[scale=0.40]{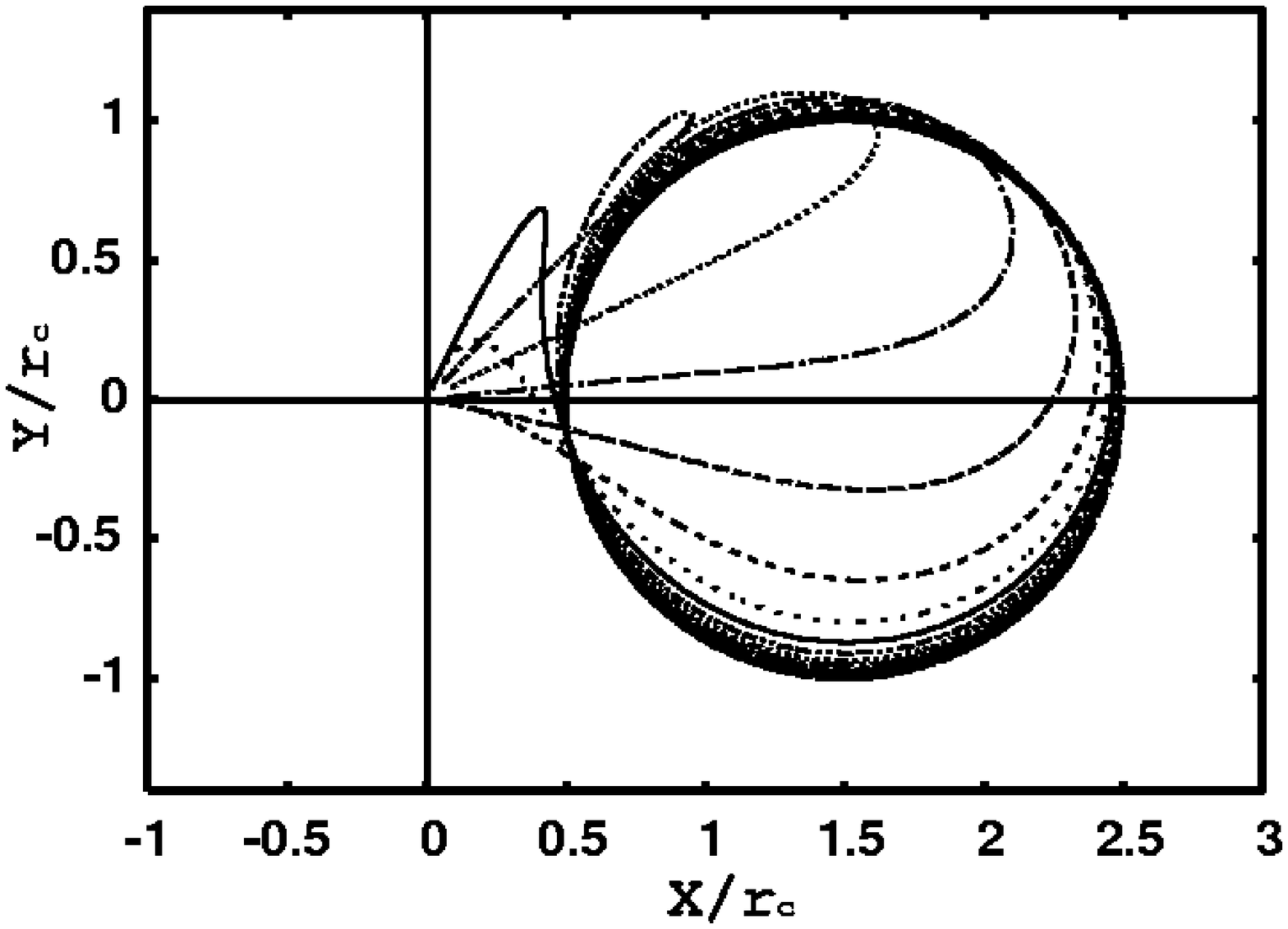}
\end{center}
\caption{
Orbits of charged particles subject to a tidal force 
are shown on the X-Y plane.
Length is normalized with respect to the Larmor radius $r_c$.
The gyro-center at the initial gyro-motion 
is located at $(r_c/2, 0)$ in the left panel 
and $(3 r_c/2, 0)$ in the right  panel.
}
\end{figure}

  In Figure 1, we demonstrate the trajectory of a charged 
particle with $ \omega >0$ experiencing a tidal force. 
The orbit at large  $ | \omega T| $ is
circular with Larmor radius $r_c$,
while the final plunging orbit depends on the phase 
in the circular motion at a certain  time.
The figure shows seven orbits whose phase is given by 
$ 2\pi n/7 ~(n=0, \cdots , 6)$ . 
The gyro-center of these orbits 
is fixed as $(r_c/2, 0)$ in the left panel, but is shifted 
to $(3 r_c/2, 0)$ in the right panel.
The tidal force causes the circular orbits to collapse, 
irrespective of the location of the gyro-center, 
and be compressed to the origin $(0, 0)$ as $T$ tends to zero.

\begin{figure}[ht]
\begin{center}
\includegraphics[scale=0.40]{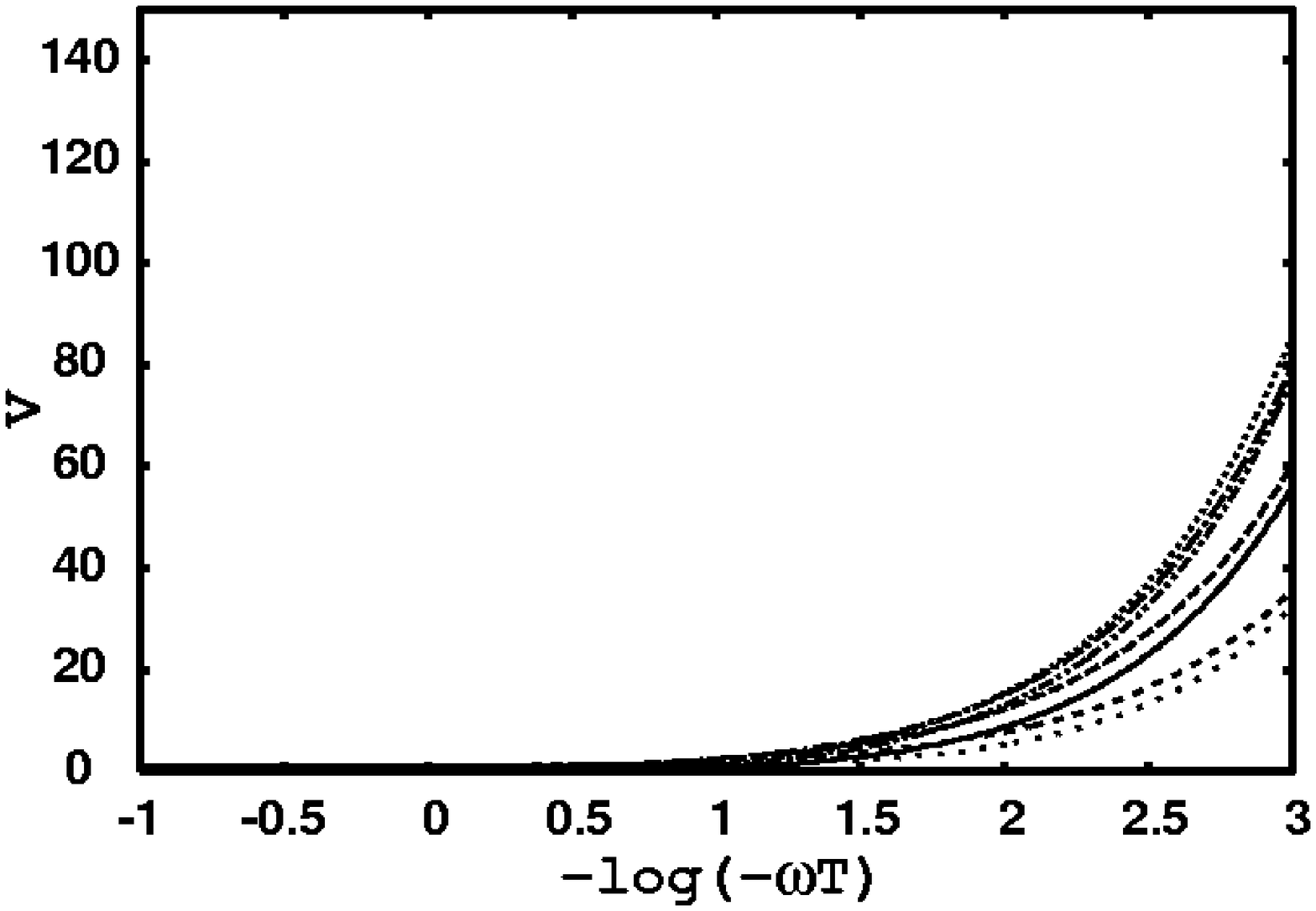}
\includegraphics[scale=0.40]{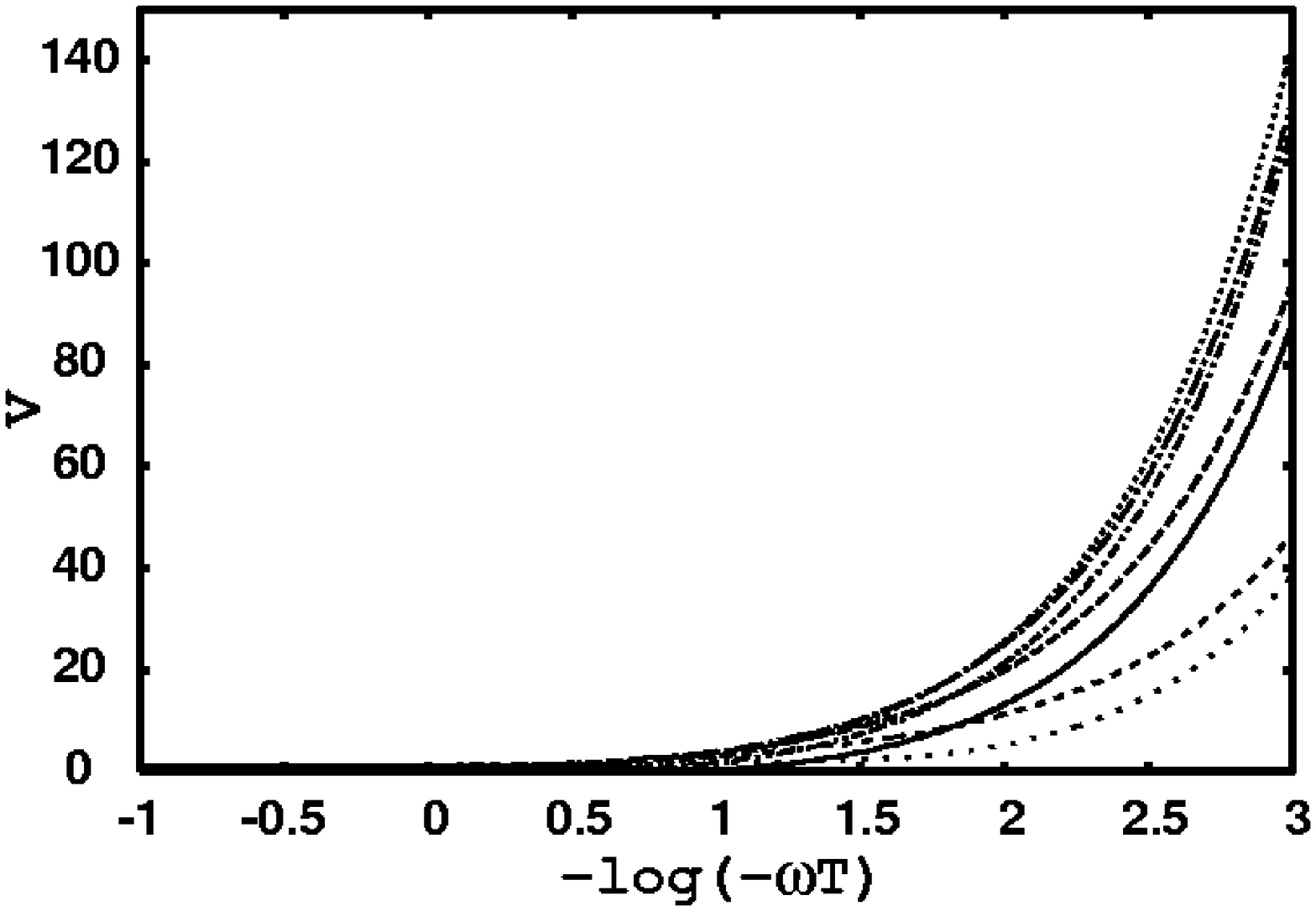}
\end{center}
\caption{
The amplification of velocity 
is shown as a function of time. 
The gyro-center of orbits is $(r_c/2, 0)$ in the left panel, but
$(3 r_c/2, 0)$ in the right panel.
}  
\end{figure}

%
The velocity $c ( \dot{X}^2 +\dot{Y}^2)^{1/2}$
is shown as a function of time in Figure 2. 
The value is normalized by the initial circular
velocity, $ v = \omega  r_c$. 
The velocity is constant during the circular motion,
but significantly increases from a certain time $T_*$
in the final plunging phase.
The transition time $T_*$ is 
$\omega T_* \approx -1 $,
obtained by  comparing 
the orders of magnitude of the Lorentz force 
$q B \omega r_c / c$ 
and the tidal force $ \sim G m M \xi/r^3 $, where
$ r (\sim (GM T^2 )^{1/3})$ is distance from 
a gravitational source with mass $M$
and the typical distance $ \xi$ from the reference observer 
is assumed to be of the same order as the Larmor radius,
$ \xi \sim r_c$.
The tidal force dominates over the Lorentz force for small $r$,
i.e, small $|T|$, and 
the condition of the tidal disruption 
can be approximated as $|\omega T |< 1 $.

  After the transition time $T_*$, the velocity increases.
The curves in Figure 2  are only shown for values of
$ \omega T $ up to $ -10 ^{-3}$, 
but actually diverge as $ \sim  |\omega T |^{2/3} $
in the limit of $\omega T \to 0  $. 
The non-relativistic treatment therefore becomes invalid,
and the special relativistic effects should be taken in account
in the final phase.

\section{Numerical results for relativistic case}

  In this section, we consider 
the tidal effect on gyration of a relativistic charged particle.
We neglect electro-magnetic radiation
from the particle for simplicity. 
The tidal acceleration is described in Fermi normal coordinates 
constructed around a reference observer.
The spatial range valid for the coordinate system 
is limited by radius $ {\cal R}^{-1/2}$, where 
$ {\cal R}$ is a typical component of curvature tensor 
evaluated along the observer's world line.
A simple criterion is
\begin{equation}
 X^2 + Y^2  < \frac{3}{2K} = \frac{27}{4} (cT)^2 .
\label{condFN}
\end{equation}
Outside this region, the spatial metric in the $X$-$Y$ plane
changes sign.

We assume that the center of the circular motion
is located not far from the reference observer at the origin,
i.e, $X \sim r_c,  Y \sim r_c $.
It is possible to realize $|X| \gg r_c,  |Y| \gg r_c $, 
but the location of the reference observer is inadequate
in such a case.
  In the non-relativistic approximation,
circular velocity $v$  in the absence of the tidal force
satisfies the relation $ v = \omega r_c $. 
Therefore, 
the location of the circular motion may be chosen 
to be close to the origin,
as long as the velocity is small enough.
Condition (\ref{condFN}) is therefore satisfied.
Formally, the right hand side of eq.(\ref{condFN}) is infinity
because the light velocity is assumed to be infinity
in non-relativistic cases.

   In relativistic cases, the circular velocity satisfies 
$ v(1 -(v/c)^2)^{-1/2} = \omega r_c $.
Therefore, $X \sim r_c,  Y \sim r_c $,  are not infinitesimal
for $ v = {\cal O}(c)  $ and a finite value of $\omega$.
 Far from the gravitational source, condition (\ref{condFN}) may be satisfied. 
As the particle falls, the curvature increases, and
 the valid spatial range accordingly decreases.
As it is difficult to know a priori whether or not
the particle always moves inside the valid spatial range, 
it is necessary to check condition (\ref{condFN})
in the evolution. 
To do this, we numerically integrate eqs.(\ref{eqx}) and (\ref{eqy})
with $Z=0$,
for relativistic orbiting particles.
Orbits of relativistic particles are calculated from large $|T|$,
provided condition (\ref{condFN}) is satisfied.
%

\begin{figure}[ht]
\begin{center}
\includegraphics[scale=0.40]{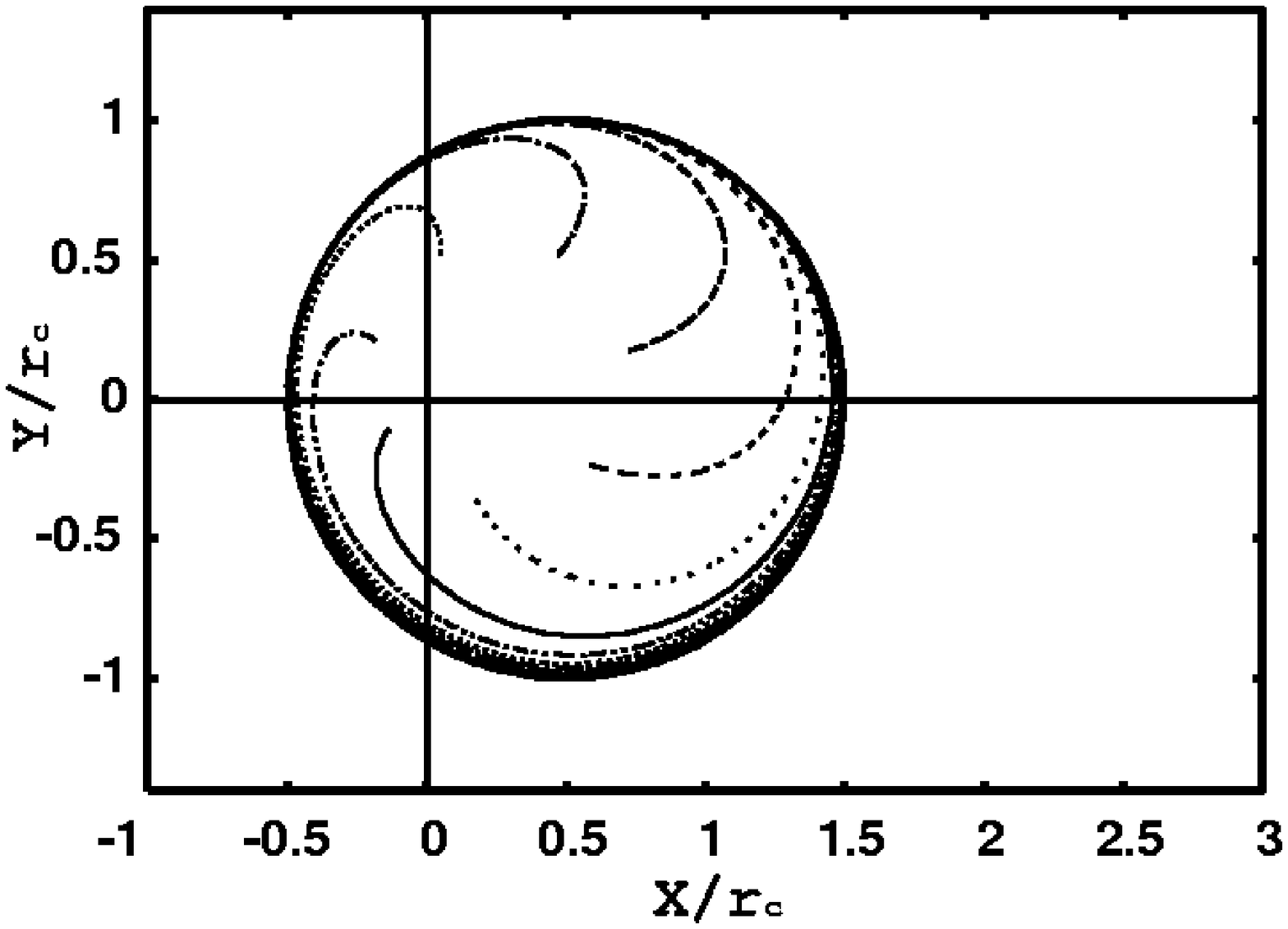}
\includegraphics[scale=0.40]{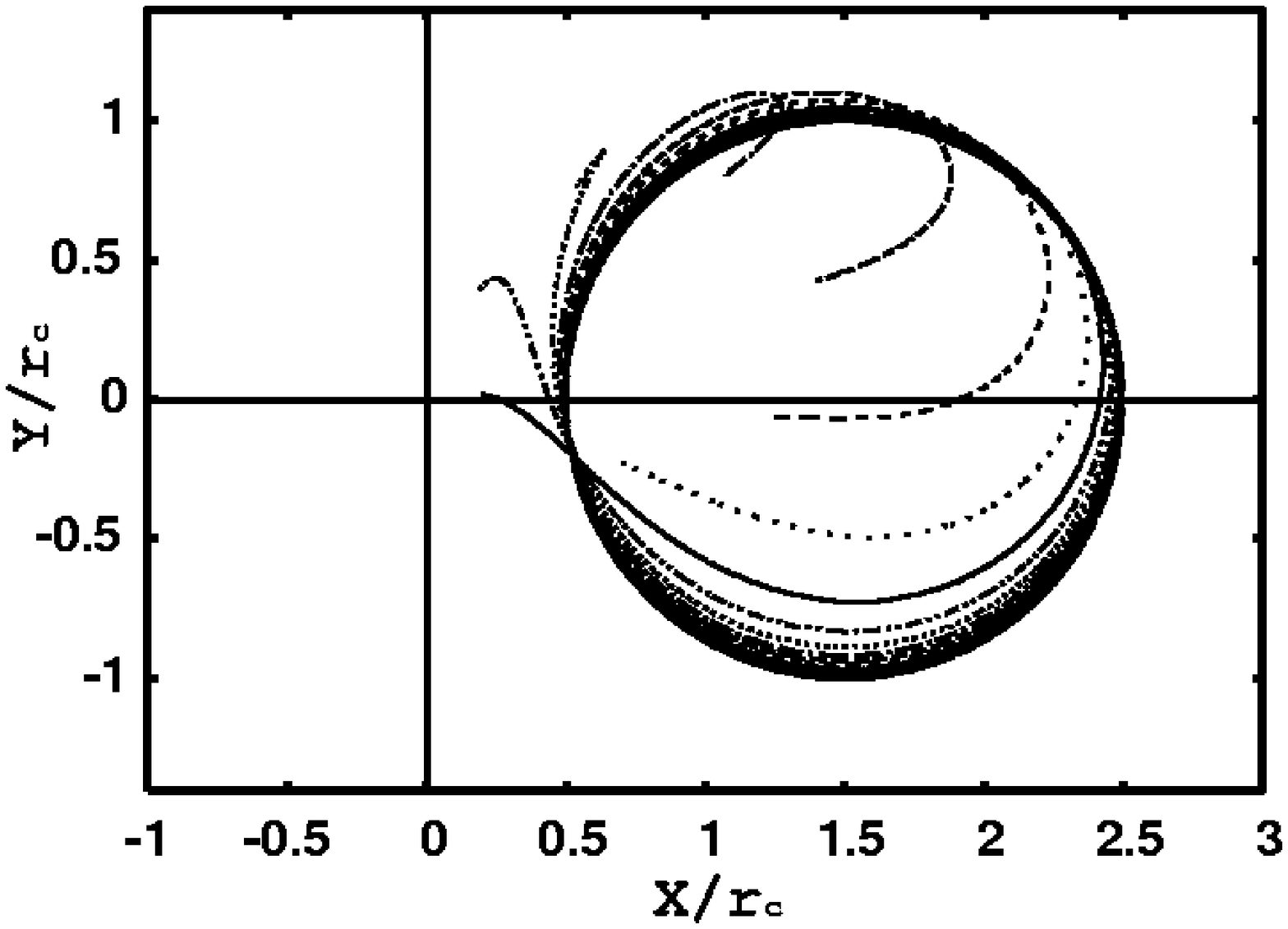}
\end{center}
\caption{
The same as in figure 1, but the initial circular velocity is
0.1c.
}
\end{figure}
\begin{figure}[ht]
\begin{center}
\includegraphics[scale=0.40]{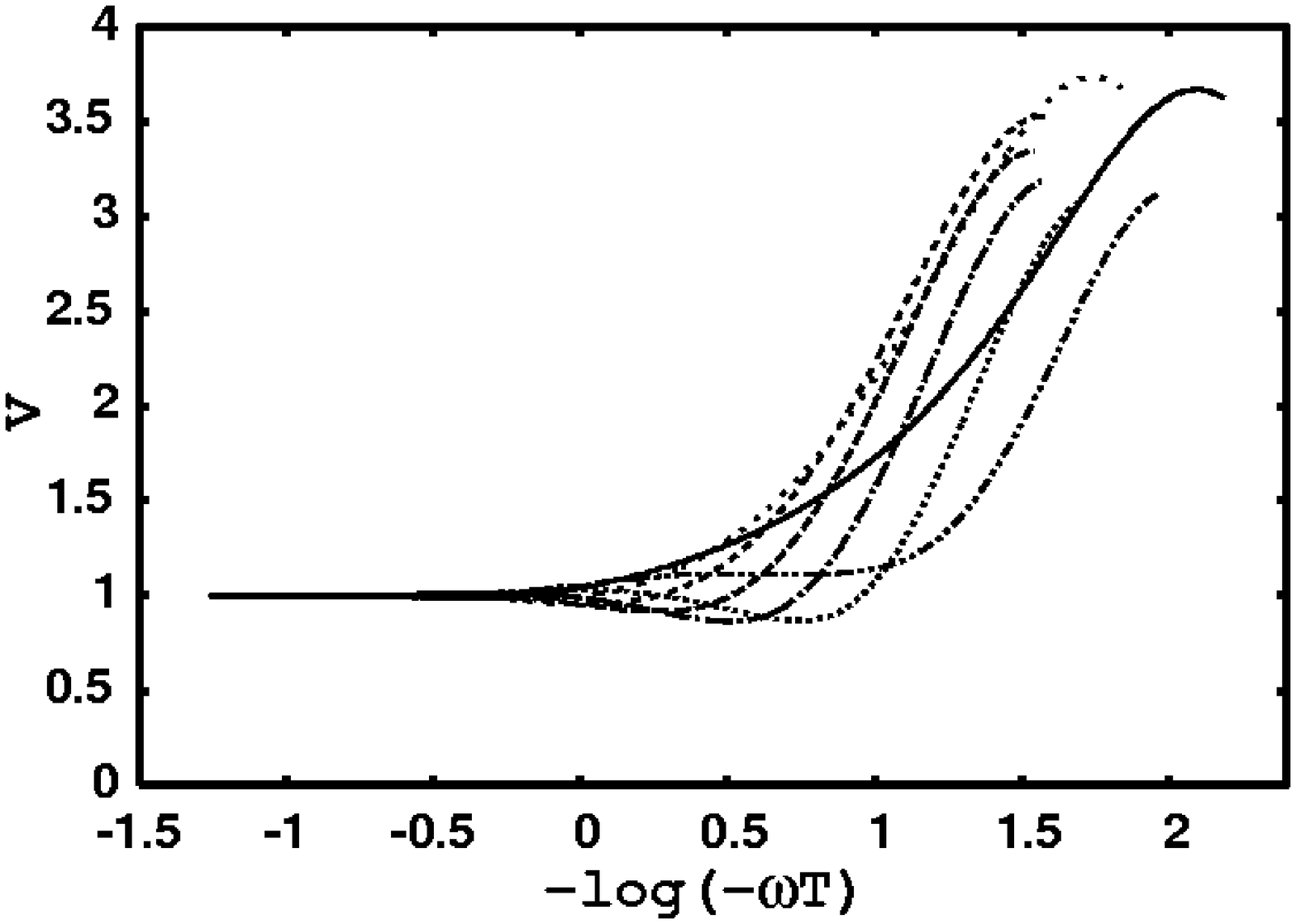}
\includegraphics[scale=0.40]{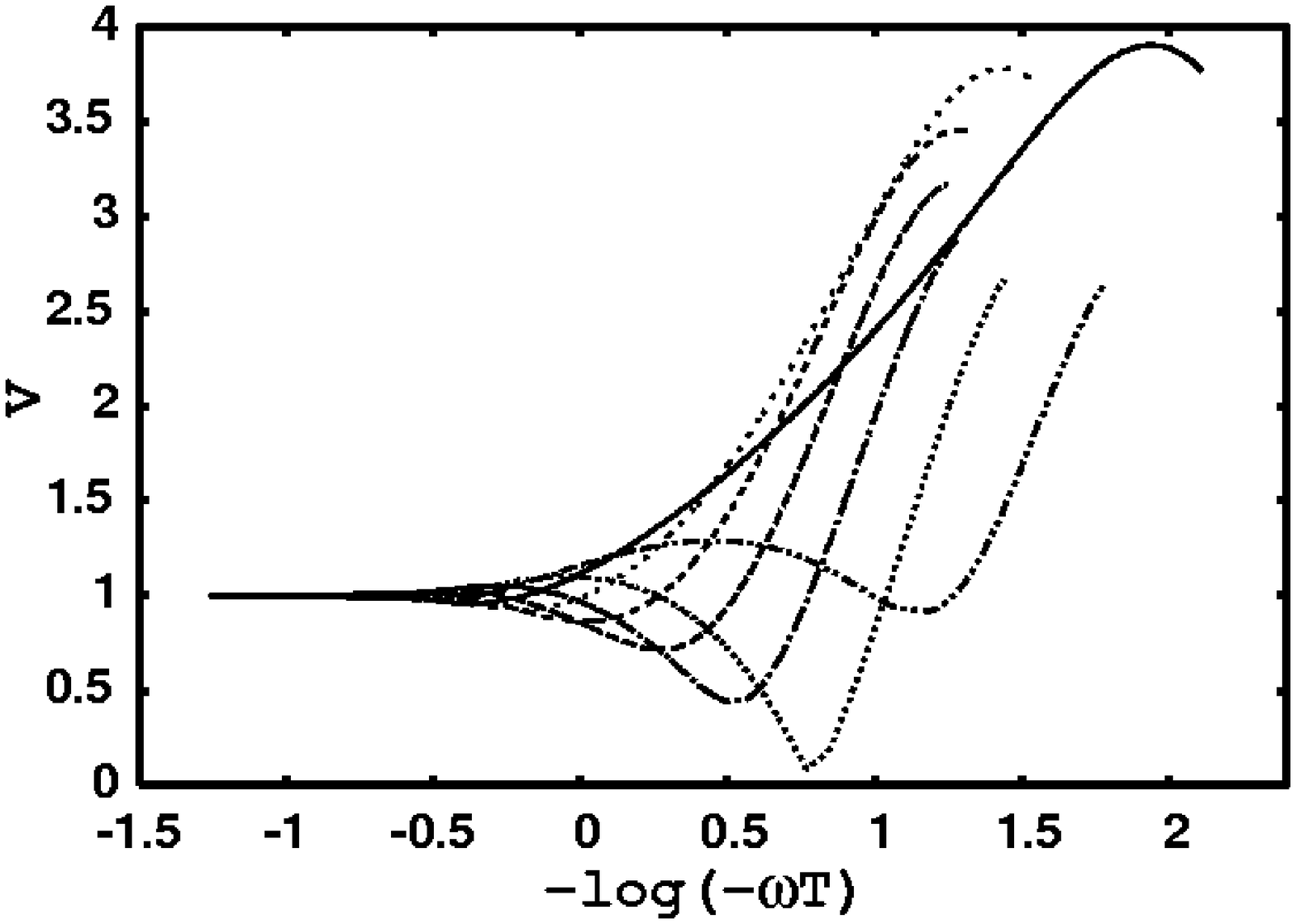}
\end{center}
\caption{
The same as in figure 2, but the initial circular velocity is
0.1c.
}  
\end{figure}

  Numerical results are given for the motion with  
an initial circular velocity $v = 0.1c$.
The trajectories are shown 
for the gyro-center $(r_c/2, 0)$ in the 
left panel and for $(3 r_c/2, 0)$ in the right panel of Figure 3.
Condition (\ref{condFN}) is no longer satisfied at the end of 
each curve, which corresponds to different time $T$.
The final phase is therefore masked because 
the particle eventually escapes outside valid 
Fermi coordinates.
This behavior seems to be general, because
the particle motion is slow compared with
the shrinkage of the valid spatial range, which is proportional 
to time $|T|$, as shown in condition (\ref{condFN}).
The orbits are almost the same as those of the non-relativistic
cases shown in Figure 1, the main difference being the 
termination points; this is due to the coordinate system.
The relativistic effects on the orbits are not apparent.

  The locally measured value of velocity $v$
at the particle position is given by
$v = c ( -g_{ij} \dot{X} ^i \dot{X} ^j /g_{00} )^{1/2}$,
which can be reduced to $c ( \dot{X}^2 +\dot{Y}^2)^{1/2}$
in flat spacetime.
The time evolution of $v$ normalized by the initial velocity
$ v=0.1c$ is shown in Figure 4.  
Like the non-relativistic cases, the velocity increases
with $ \omega T$ 
from the transition time $\omega T_* \approx -1 $.
 However, the final behavior, i.e,
divergence of velocity seen 
in non-relativistic cases 
cannot be treated in this system, 
because the particle position is outside the 
valid coordinate system.
The sudden drop in velocity in the right panel of Figure 4 
corresponds to the kink in the orbit shown in  Figure 3.
This is caused by the tidal acceleration almost reversing 
the direction of the particle's orbit.

\begin{figure}[ht]
\begin{center}
\includegraphics[scale=0.40]{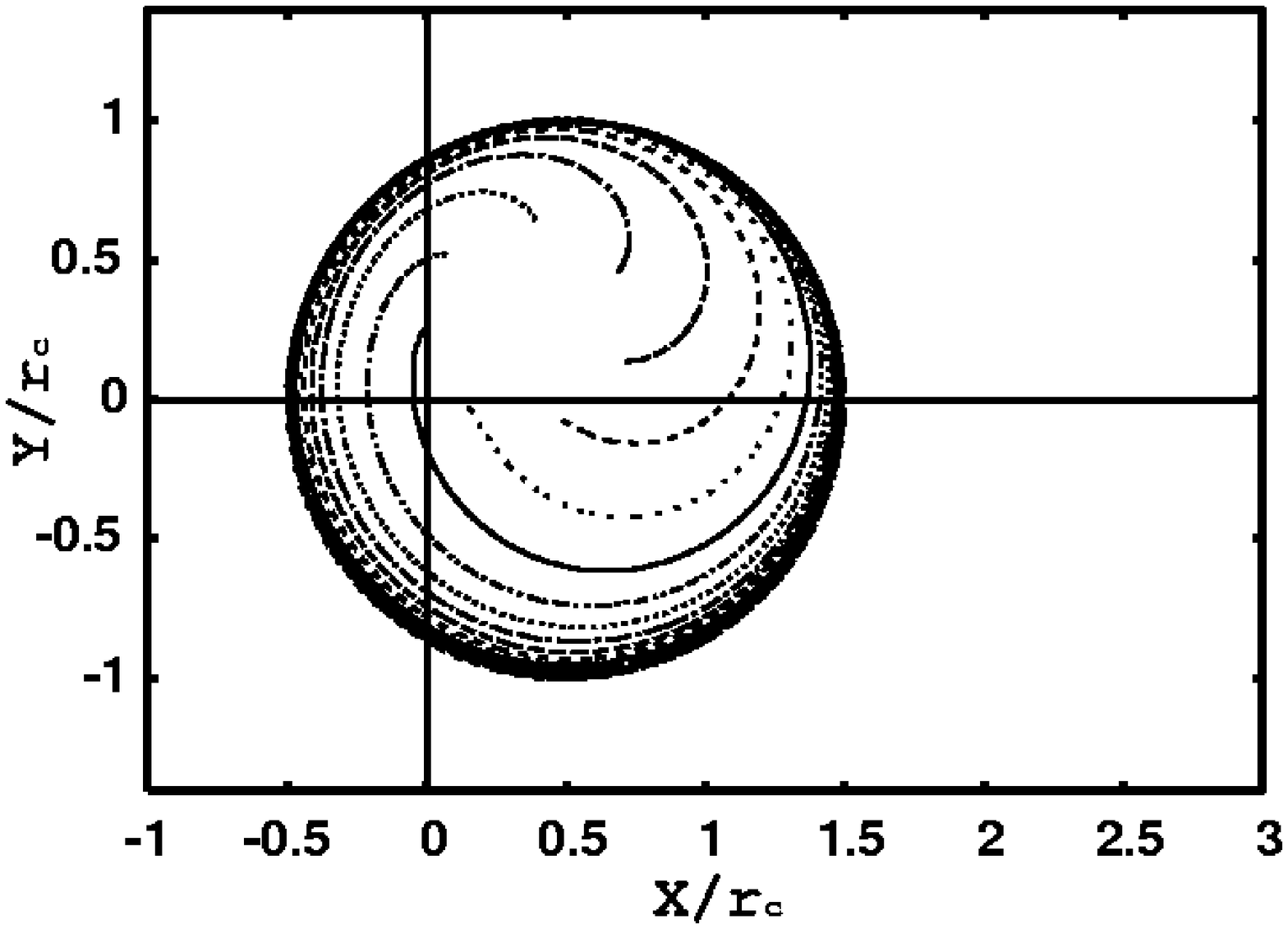}
\includegraphics[scale=0.40]{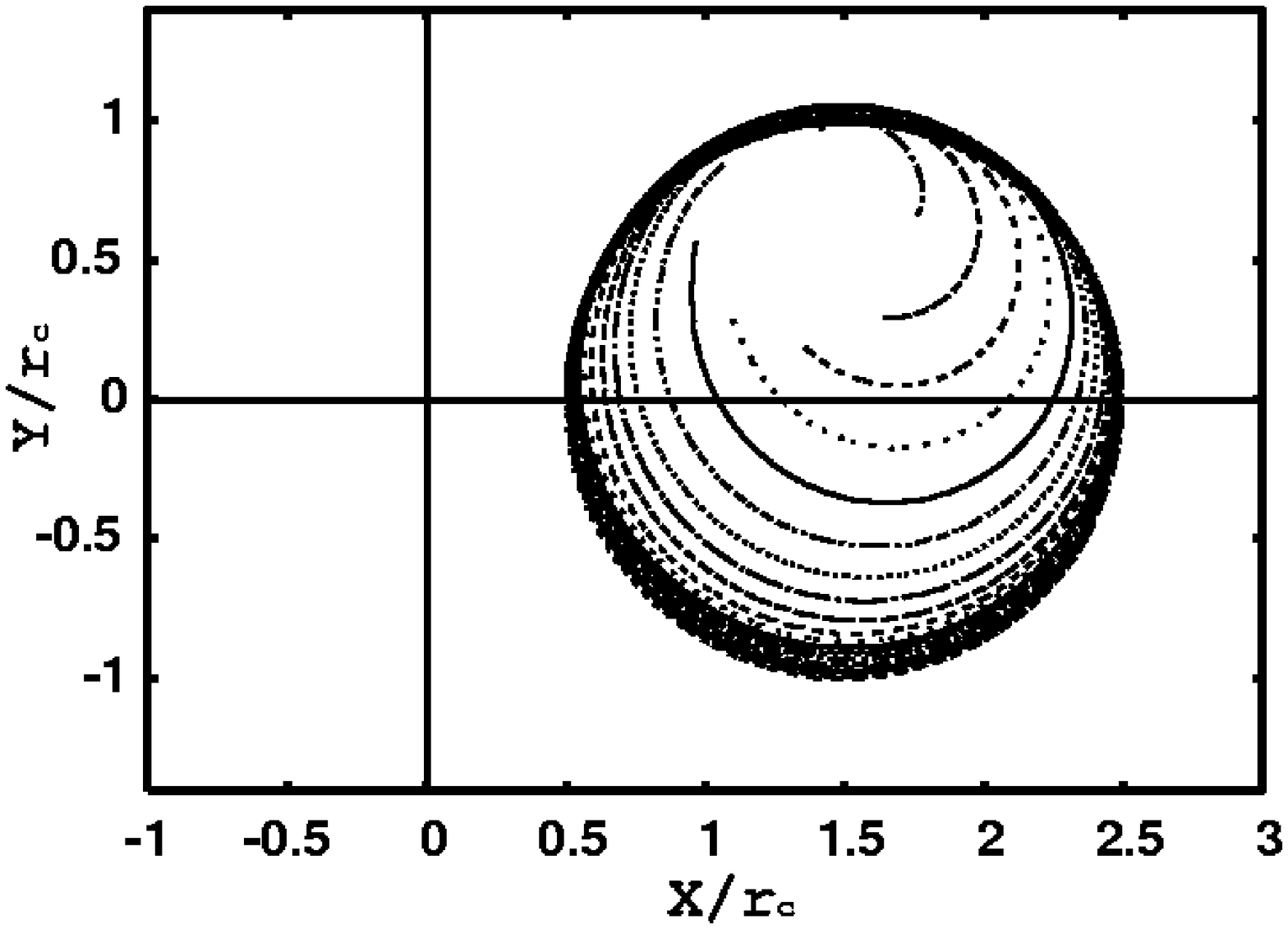}
\end{center}
\caption{
The same as in figure 1, but the initial circular velocity is
0.9c.}  
\end{figure}
\begin{figure}[ht]
\begin{center}
\includegraphics[scale=0.40]{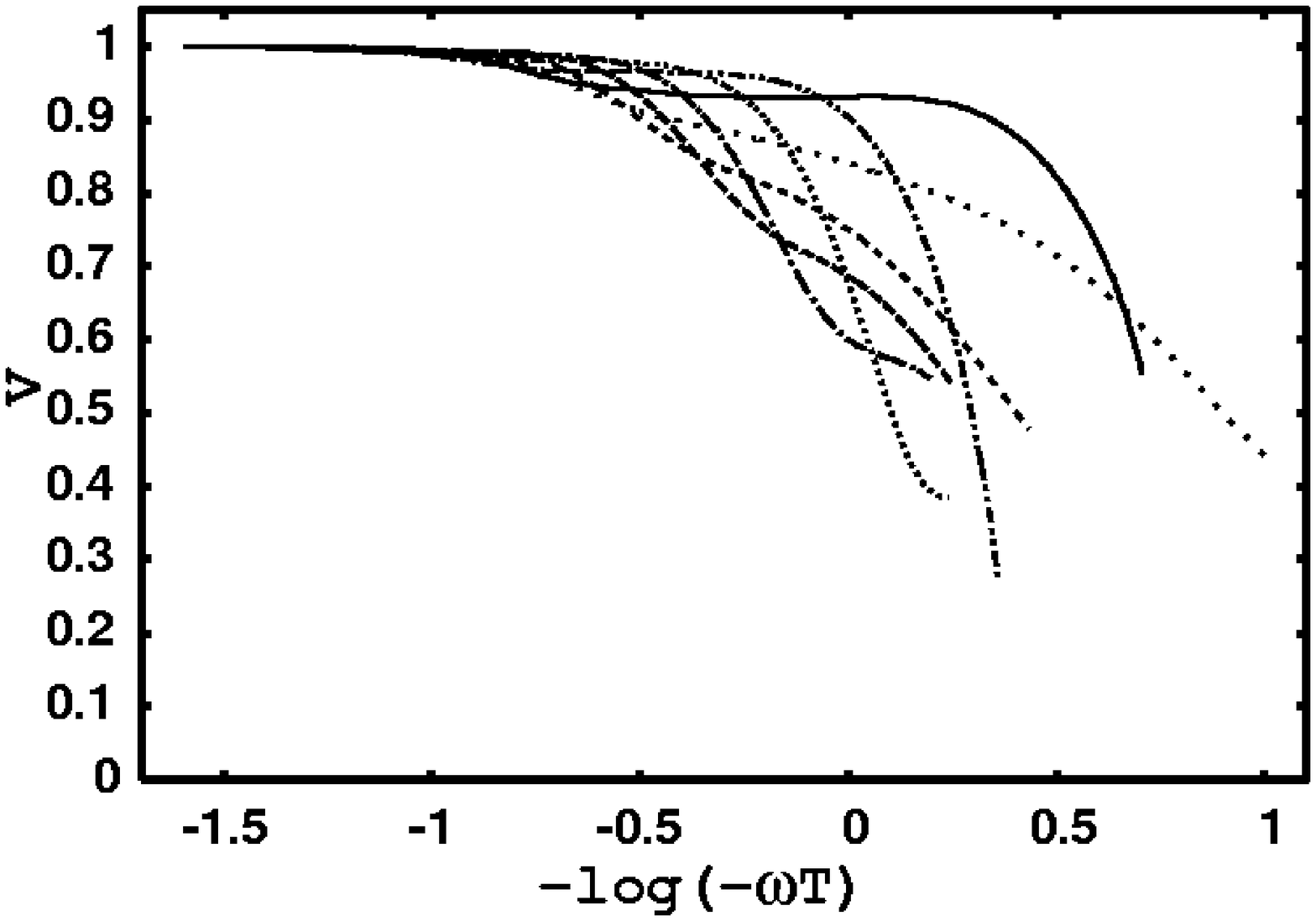}
\includegraphics[scale=0.40]{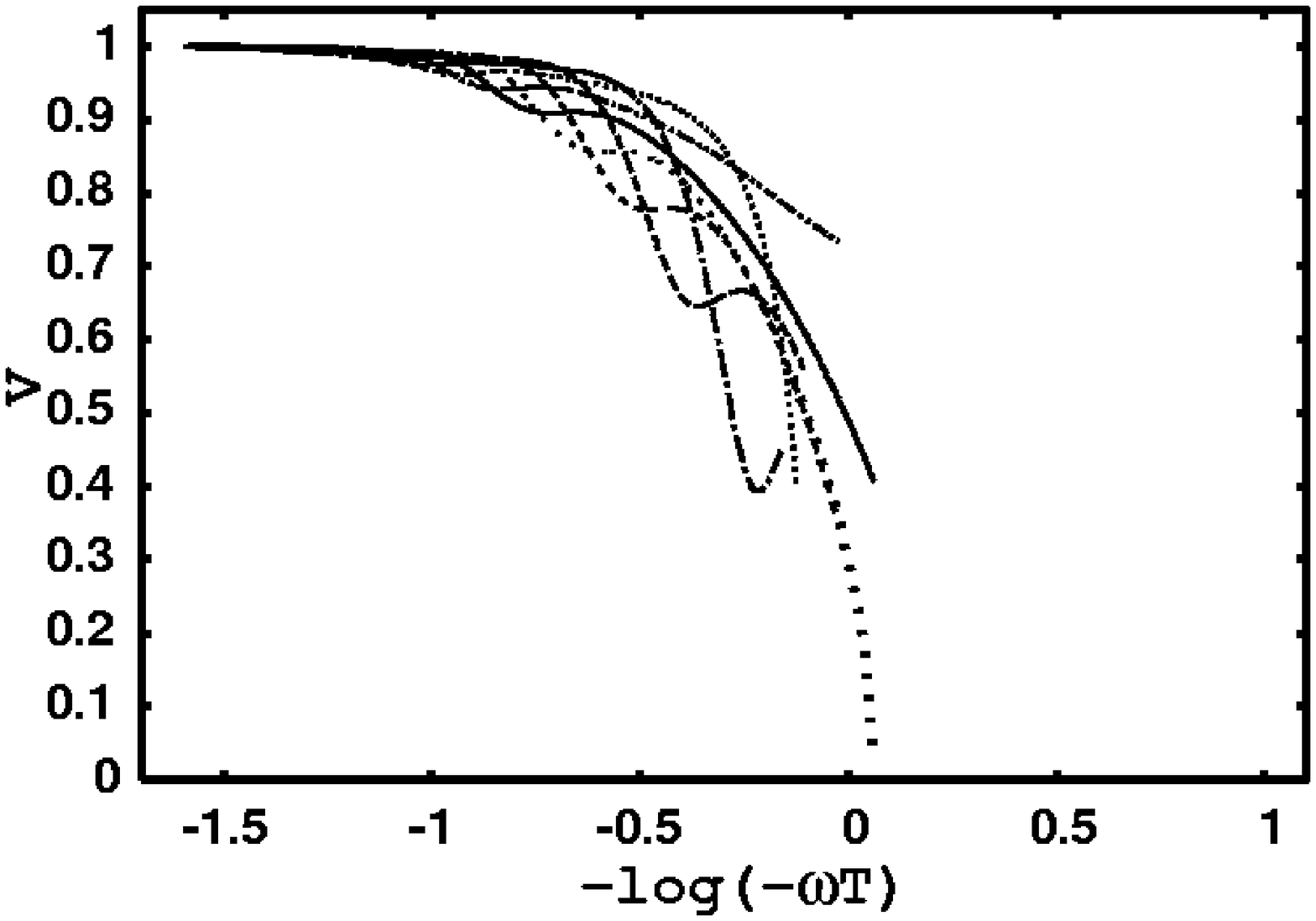}
\end{center}
\caption{
The same as in figure 2, but the initial circular velocity is
0.9c.}  
\end{figure}
%
It is very interesting to consider the tidal effect on
a relativistic particle with a much larger initial velocity.
Is the particle accelerated up to the velocity of light? 
We show the results for initial circular velocity, $v = 0.9c $.
The orbits are shown for gyro-center $(r_c/2, 0)$ in the
left panel and for $(3 r_c/2, 0)$ in the right panel of Figure 5.
The spiral of the orbits is much tighter due to
the larger circular velocity. 
In their final phase, the particles always escape from the 
valid coordinate system, as in Figure 3,
preventing us from following the subsequent evolution.
This termination may artificially cause
these orbits to converge to 
$( 0.5 r_c , 0.2 r_c )$ in the left panel and
$( 1.5 r_c , 0.5 r_c )$ in the right  panel.
Each position suggested from the figure
is not a true convergent position because
the end points do not  correspond to the same
coordinate time.
Curves that stopped at early 
have larger $ (X^2 +Y^2)^{1/2}$ due to  condition (\ref{condFN}).

  The time evolution of velocity
$v = c ( -g_{ij} \dot{X} ^i \dot{X} ^j /g_{00} )^{1/2}$
is shown in Figure 6.
It is normalized by the initial velocity $ v=0.9c$.
Remarkably, the velocity never increases, and actually decreases.
In this case, the tidal force causes deceleration,
contrary to our intuition.
The physical mechanism is not easily understood, but
the mathematical origin  is simple.
The tidal acceleration terms coupled with velocity
in eqs.(\ref{eqx}) and (\ref{eqy})
change sign for large velocities of
$ \dot{X} \sim \dot{Y} \sim {\cal O} (1) $.
Beyond a critical value, the tidal force becomes
repulsive in the relativistic cases, 
whereas it is attractive in the non-relativistic cases. 
This property is analogous to that
reported by Chicone and Mashhoon \cite{CM04}.
We have calculated evolution for various values of
the initial circular velocity in order 
to determine whether the tidal force 
causes acceleration or deceleration. 
It was found that the critical initial velocity is approximately
$ \sim 0.7c$, and that the velocity decreases 
above this value.

\section{Summary and discussion }

  In this paper, we have examined the tidal effect on 
a charged particle orbiting in a uniform magnetic field.
From this simple toy model, we found that the tidal force 
changes the property when velocity is relativistic, 
as shown by Chicone and Mashhoon \cite{CM04}.
In our model, the gyration velocity on the $X$-$Y$
plane changes due to the tidal force coupled to the velocity,
as the particle falls into a black hole along the $Z$ axis.
The velocity at the final plunging phase increases,
when the initial velocity is less than a critical value. 
On the other hand, the velocity decreases, 
when it is larger than the critical value. 
The critical value was about $0.7c$
in our model, but depends on the process.

  As for speculative effects of the velocity-dependent
tidal force in a more realistic astrophysical situation,
suppose that a spinning neutron star falls into
a black hole along the $Z$ axis.
The property of the tidal force depends 
on the rotating velocity 
$ v_\phi $, which corresponds to the orbiting velocity
of the particle in our model.
When $ v_\phi $ exceeds a certain value,
the tidal disruption may occur in 
a peculiar way, quite different from that in 
non-relativistic cases.
The maximum rotation velocity of a neutron star is 
of the order $v_\phi \sim  0.5c$ 
at the  equatorial plane (e.g, \cite{gle}).
Therefore, it is very interesting 
to investigate the evolutionary behavior and the critical 
velocity in the hydrodynamical process with
the velocity-dependent tidal force.

 As for the valid range of 
the coordinate system,
the Fermi normal coordinates are set up around a 
geodesic reference observer,
we used second-order expansion for the
metric about the reference observer, and neglected
higher order terms.
The tidal force appears due to non-uniformities
in the gravitational field, and is expressed by 
the deviation from the flat spacetime.
The force becomes significant when the curvature tensors
measured by the reference observer are large.
At the same time, the higher order terms are no longer
negligible.
In  other words, the valid range of the coordinate 
system shrinks as the gravitational source is approached.
In order to study the final phase of the tidal disruption,
it is important to include the  
higher order expansion terms.
See e.g., \cite{LN79,ISM05} for 
the third order and fourth order expansion terms.

\ack
This work was supported in part 
by a Grant-in-Aid for Scientific Research 
(No.16029207 and No.16540256) from 
the Japanese Ministry of Education, Culture, Sports,
Science and Technology.
\section*{References}

   \end{document}